\documentclass[11pt,a4paper]{article}
\RequirePackage{amsmath,amssymb}
\RequirePackage[dvipsnames,usenames]{color}
\usepackage{cite}
\usepackage{fullpage}
\usepackage[british]{babel}
\usepackage[latin1]{inputenc}
\usepackage[T1]{fontenc}
\usepackage[final]{showkeys} 
\usepackage[bookmarks]{hyperref}
\usepackage{amsthm}
\usepackage{graphicx}
\usepackage{subfigure}
\usepackage{braket}
\usepackage{mathrsfs}

\usepackage{bm}
\usepackage{amsfonts}
\usepackage{braket}

\newcommand{\bc}{\begin{center}}
\newcommand{\ec}{\end{center}}

\newcommand{\be}{\begin{eqnarray}}
\newcommand{\ee}{\end{eqnarray}}

\title{\bf Revisiting the conformal invariance of Maxwell's equations in 
curved spacetime}

\author{
Jeremy~C\^ot\'e\thanks{E-mail: jcote16@ubishops.ca},
Valerio~Faraoni\thanks{E-mail: vfaraoni@ubishops.ca},
~and
Andrea~Giusti\thanks{E-mail: agiusti@ubishops.ca}     
$\,$
\\
\\
{\em Department of Physics and Astronomy, Bishop's University}
\\
{\em 2600 College Street, Sherbrooke Qu\'ebec, Canada J1M 1Z7}
}

\begin{document}
\maketitle
\begin{abstract}

We revisit the invariance of the curved spacetime Maxwell equations under 
conformal transformations. Contrary to standard 
literature, we include the discussion of the four-current, the wave 
equations for the four-potential and the field, and the behaviour of gauge 
conditions under the conformal transformation. 

\end{abstract}

\newpage

\section{Introduction}
\label{sec:1}
\setcounter{equation}{0}

Conformal transformations of the spacetime metric constitute a very useful 
tool of general relativity \cite{Penrose, Penrose2, Penrose3, Synge, MTW, 
Wald, Carroll}. Alternative theories of gravity use conformal 
transformations even more heavily: for example scalar-tensor gravity 
admits two representations related by a conformal transformation, the 
so-called Jordan and Einstein conformal frames \cite{Dicke, Harrison}. The 
conformal transformations we refer to should not be confused with the 
coordinate transformations of the conformal group in flat space, which 
also leave the Maxwell equations invariant \cite{Cunningham, Bateman, 
Weyl, Fulton}.

A {\em conformal transformation} is a position-dependent rescaling of the 
spacetime metric
\be 
g_{ab} \rightarrow \tilde{g}_{ab}=\Omega^2 g_{ab}  \label{conftransf}
\ee
where the {\em conformal factor} $\Omega( x^{\alpha}) $ is a 
(dimensionless) 
positive smooth function of the spacetime position $x^{\alpha}$. Conformal 
transformations do not 
change the metric signature, the sign of the magnitude of 
four-vectors, the angles between them and, more important, they leave 
the light cones and the causal structure of spacetime invariant 
(see Appendix~D of Ref.~\cite{Wald}).  

It is standard knowledge that the Maxwell equations in four 
spacetime dimensions  are invariant under conformal transformations 
\cite{Synge, MTW, Wald, Carroll}. The physical interpretation of this fact 
is 
that, due to the 
fact that the photon is massless, no length or mass scale 
is associated with the electromagnetic field.\footnote{By contrast, the 
equations for the Proca field, which contain a mass scale, are not 
conformally invariant.} Therefore, the Maxwell 
equations are not affected by a (point-dependent) rescaling of the 
metric which changes non-zero distances between points and the lengths of 
non-null 
vectors. In the geometric optics limit of wavelengths negligible in 
comparison with the radius of curvature of spacetime \cite{Wald, 
Carroll, Stephani}, 
electromagnetic waves travel along null geodesics and it is well 
known that a conformal 
transformation~(\ref{conftransf}) leaves null geodesics invariant, apart 
from a change of parametrization \cite{Wald, Carroll}. 

The standard proof of the conformal invariance of the 
Maxwell equations ({\em e.g.}, Appendix D of Ref.~\cite{Wald}) is presented in the absence of 
sources of the  electromagnetic field and it refers to the 
equations satisfied by the Maxwell tensor $F_{ab}$,
\begin{eqnarray}
&& \nabla^b F_{ab}=0 \,,\label{Maxwell1}\\
&&\nonumber\\
&& \nabla_{[a}F_{bc]}=0 \,,\label{Maxwell2}
\end{eqnarray}
where $\nabla_c$ denotes the covariant derivative operator of the 
spacetime metric 
$g_{ab}$, square brackets around indices denote antisymmetrization, we 
follow the notation of Ref.~\cite{Wald}, and we restrict to four 
spacetime dimensions (this last assumption is crucial for the conformal 
invariance of the Maxwell equations: in higher dimension, conformal 
invariance can be achieved only at the price of modifying substantially 
the Maxwell action \cite{Deser}). Nothing is usually said about the 
conformal 
invariance of the equation satisfied by the electromagnetic 
four-potential $A_b$ which, in the absence of sources, is  
\be
\Box A_b -\nabla_b \nabla^c A_c -{R_b}^d A_d=0 \,, 
\ee
where $\Box \equiv g^{ab} \nabla_a \nabla_b $ 
is d'Alembert's operator in curved spacetime. Indeed, the standard 
presentation of this equation is in the Lorentz gauge $\nabla^c A_c=0$, in 
which the term $-\nabla_b \nabla^c A_c$ drops out. While the physical 
justification for the conformal invariance of the source-free Maxwell 
equations~(\ref{Maxwell1}), (\ref{Maxwell2}) is intuitive, three questions 
arise.

First, are the Maxwell equations in the presence of charges and currents 
(described by the four-vector $j^a$) 
\begin{eqnarray}
&& \nabla^a F_{ab}= -4\pi j_b \,,\label{Maxwells1}\\
&&\nonumber\\
&& \nabla_{[a}F_{bc]}=0 \,,\label{Maxwells2}
\end{eqnarray}
still conformally invariant?

The answer is not trivial because, while 
the Maxwell field is associated with the massless photon, with the 
exception of displacement currents, charges and 
currents are associated with matter (electrons, protons, or ions). The 
answer is that the Maxwell 
equations with sources are still conformally invariant, but 
apparently the proof does not appear in the literature. It is presented here, 
together with the scaling property of the four-current $j^a$ under 
conformal transformations and with the explicit verification 
of the conformal invariance of charge conservation.

Second, is the Maxwell equation satisfied by the 
electromagnetic four-potential 
\be
\Box A_b -\nabla_b \nabla^c A_c -{R_b}^d A_d= -4\pi j_b 
\label{full}
\ee
(written here in the presence of sources)
still conformally invariant?

After all, there are substantial 
differences between Eq.~(\ref{full}) and the Maxwell 
equations~(\ref{Maxwells1}), (\ref{Maxwells2}): (\ref{full}) is a wave 
equation while (\ref{Maxwells1}) and (\ref{Maxwells2}) are first order 
equations for the field $F_{ab}$. Further,  $A^b$ 
couples explicitly to the Ricci tensor. One might wonder what happens to 
$A^c$, since it is gauge-dependent. Under a conformal transformation, it 
seems plausible that the conformal invariance is broken. However, if 
these terms that break the conformal invariance are pure gauge terms 
(which give a vanishing contribution to $F_{ab}$), then the situation 
might not be as bad. It turns out (but is not usually mentioned in the 
literature) that Eq.~(\ref{full}) is conformally invariant and, like the 
Maxwell field $F_{ab}$, the four-potential $A_b$ is conformally 
invariant.

Third, given that the equation for $A^b$ is usually presented in the 
Lorentz gauge $\nabla^c A_c=0$, is this gauge (or any gauge choice) 
preserved by a conformal transformation? In general, the answer is 
negative, as will be shown in Sec.~\ref{sec:3Lorentzgauge}.

Before computing the answers to the questions above, we recall the action 
for the electromagnetic field with sources described by the four-current 
$j^a$ in curved spacetime \cite{MTW, Stephani, Wald, Carroll}
\be
S_{(em)} =\int d^4 x \sqrt{-g} \, \left( -\frac{1}{4} \, F_{ab}F^{ab} 
+4\pi 
A_b j^b \right) = 
\int d^4 x \sqrt{-g} \, \left( - g^{ac} g^{bd} \nabla_{[a} A_{b]} 
\nabla_{[c}  A_{d]} + 4\pi A_b j^b \right) \,, \label{emaction}
\ee
where $g$ is the determinant of the metric tensor $g_{ab}$. The variation 
of the action~(\ref{emaction}) with respect to $g^{ab}$  and $A^c$ 
produces the field equations~(\ref{Maxwells1}) and (\ref{Maxwells2}), 
respectively ({\em e.g.}, Ref.~\cite{Carroll}, p.~164). The 
stress-energy tensor of the electromagnetic field is 
\be
T_{ab}= -\, \frac{2}{\sqrt{-g}} \, \frac{\delta 
S_{(em)} }{\delta g^{ab}} = \frac{1}{4\pi} 
\left( F_{ac} {F_b}^c -\frac{1}{4} \, g_{ab}  
F_{de} F^{de} \right) \,, \label{Tab}
\ee
has vanishing trace $T\equiv g^{ab}T_{ab}=0$, and it is not covariantly 
conserved in the presence of sources $j^c$  
interacting with the field and exchanging energy and momentum with it. By 
taking the covariant divergence of $T_{ab}$ and using the field 
equations~(\ref{Maxwells1}) and (\ref{Maxwells2}), one easily obtains 
\be
\nabla^b T_{ab} = - F_{ab} j^b \,.
\ee
We also need the transformation properties  
of various spacetime quantities  under conformal 
rescalings \cite{Synge, Wald, Carroll}, including the inverse metric
\be
\tilde{g}^{ab} = \Omega^{-2} g^{ab} \,, \label{eq:t1}
\ee
\be
\sqrt{-\tilde{g}} = \Omega^4 \sqrt{-g} \,, \label{eq:t2}
\ee
the Christoffel symbols 
\be
\tilde{\Gamma}^a_{bc} = \Gamma^a_{bc} +\Omega^{-1} \left( 
\delta^a_b \nabla_c \Omega + 
\delta^a_c \nabla_b \Omega -
g_{bc} \nabla^a \Omega \right) \,, \label{Gammascaling}
\ee
and the Ricci tensor
\begin{eqnarray}
\tilde{R}_{ab} &=& R_{ab}- 2 \nabla_a \nabla_b \ln \Omega -g_{ab} g^{ef} 
\nabla_e \nabla_f \ln \Omega
+ 2 \nabla_a \ln\Omega \nabla_b \ln\Omega  -2g_{ab} \, g^{ef} \nabla_e 
\ln\Omega \nabla_f \ln\Omega \,. \nonumber\\
&&    \label{eq:t4}
\end{eqnarray}

\section{Maxwell equations with sources}
\label{sec:2}

Let us consider the Maxwell equations~(\ref{Maxwells1}), 
(\ref{Maxwells2}) with sources described by the  
four-current 
$j^a$. Under the conformal transformation~(\ref{conftransf}), the Maxwell 
tensor $F_{ab}$ will scale\cite{Wald} with conformal weight $s$,
\be
\tilde{F}_{ab} = \Omega^s F_{ab} \,,
\ee
where a tilde denotes quantities in the conformally rescaled spacetime 
with metric  $\tilde{g}_{ab}$ \cite{Wald}. The covariant divergence of the 
Maxwell tensor (needed in the first Maxwell equation) in this rescaled 
world is
\begin{eqnarray}
\tilde{g}^{ac}  \tilde{\nabla}_c \tilde{F}_{ab} &=& \Omega^{-2} g^{ac} 
\tilde{\nabla}_c \left( \Omega^s F_{ab} \right) \nonumber\\
&&\nonumber\\
&=& \Omega^{s-2} g^{ac} \tilde{\nabla} F_{ab} + s \Omega^{s-3} g^{ac} 
F_{ab} \nabla_c \Omega \nonumber\\
&&\nonumber\\
&=& \Omega^{s-2} g^{ac} \left( \partial_c F_{ab}
- \tilde{\Gamma}^e_{ca} F_{eb} 
- \tilde{\Gamma}^e_{cb} F_{ae} \right)
+s\Omega^{s-3} g^{ac} F_{ab} \nabla_c \Omega \nonumber\\   
&&\nonumber\\
& = & \Omega^{s-2} g^{ac} \left[ \partial_c F_{ab}
- \Gamma^e_{ca} F_{eb}
- \Gamma^e_{cb} F_{ae} 
-\Omega^{-1} \left( 
\delta^e_c \nabla_a \Omega + \delta^e_a \nabla_c \Omega  - g_{ac} \nabla^e 
\Omega   \right) F_{eb} \right. \nonumber\\
&&\nonumber\\
&\, & \left. -\Omega^{-1} \left( 
\delta^e_c \nabla_b \Omega + \delta^e_b \nabla_c \Omega  - g_{bc} \nabla^e 
\Omega   \right) F_{ae} \right]
+s\Omega^{s-3} g^{ac} F_{ab} \nabla_c \Omega \\
&&\nonumber\\
& = & \Omega^{s-2} \nabla^a F_{ab} -s \Omega^{s-3} F_{be}\nabla^e \Omega 
\,. 
\end{eqnarray}
Using the Maxwell equation~(\ref{Maxwells1}), one obtains
\be
\tilde{g}^{ac} \tilde{\nabla}_c \tilde{F}_{ab} = -4\pi \Omega^{s-2} j_b -s 
\Omega^{s-3} F_{be} \nabla^e \Omega \,,
\ee
while the left hand side of the second Maxwell 
equation~(\ref{Maxwells2}) in the rescaled world is
\be
\tilde{\nabla}_{[a} \left( \Omega^s F_{bc]} \right) = \Omega^s 
\nabla_{[a}F_{bc]} +s \Omega^{s-1} \left( \nabla_{[a} \Omega \right) 
F_{bc]} \,;
\ee
it is clear that the only value of the conformal weight of $F_{ab}$ that 
leaves both Maxwell equations conformally invariant in the rescaled 
spacetime is $s=0$. Using this value, the 
electromagnetic tensor with two covariant indices is conformally 
invariant,
\be
\tilde{F}_{ab} = F_{ab} \,, \;\;\;\;\;
\tilde{F}^{ab} = \Omega^{-4} F^{ab} \,, \;\;\;\;\;
 {\tilde{F}_a  }^{\,\,b} = \Omega^{-2} {F_a}^b \,, \label{eq:20}
\ee
and the conformally rescaled Maxwell equations read
\begin{eqnarray}
&& \tilde{g}^{ac} \tilde{\nabla}_c \tilde{F}_{ab}= -4\pi \tilde{j}_b 
\,,\\
&&\nonumber\\
&& \tilde{\nabla}_{[a} \tilde{F}_{bc]}=0 \,,
\end{eqnarray}
provided that the four-current transforms according to  
\be
\tilde{j}_b = \Omega^{-2} j_b  \label{tildej}
\ee
(we are not aware of occurrences of this last equation in the literature).

\subsection{Covariant charge conservation}

One can now check explicitly that the electric charge is covariantly 
conserved 
in the conformally rescaled geometry. The covariant divergence 
(according to the 
metric $\tilde{g}_{ab}$) of the rescaled four-current $\tilde{j}^b$ 
given by Eq.~(\ref{tildej}) is
\begin{eqnarray}
\tilde{g}^{ac} \tilde{\nabla}_a \tilde{j}_c &=& \Omega^{-2} g^{ac} 
\tilde{\nabla}_a \left( \Omega^{-2}  j_c \right) \nonumber\\
&&\nonumber\\
&=& \Omega^{-2} g^{ac} \left[ -2 \Omega^{-3} \left( \nabla_a \Omega 
\right) j_c +\Omega^{-2} \tilde{\nabla}_a j_c \right] \nonumber\\
&&\nonumber\\
&=& \Omega^{-2} g^{ac} \left[ -2 \Omega^{-3} \left( \nabla_a \Omega
\right) j_c + \Omega^{-2} \left( \partial_a j_c - \tilde{\Gamma}^d_{ac} 
\, j_d \right) \right] \nonumber\\
&&\nonumber\\
&=& \Omega^{-2} g^{ac} \left\{ 
-2 \Omega^{-3} \left( \nabla_a \Omega \right) j_c 
+ \Omega^{-2} \left[   \partial_a j_c - \Gamma^d_{ac} \, j_d 
\right.\right.\nonumber\\
&&\nonumber\\
&\, & \left.\left. - \Omega^{-1} \left( 
\delta^d_a \nabla_c \Omega + \delta^d_c \nabla_a \Omega -g_{ac} 
\nabla^d \Omega \right) j_d \right] \right\} \nonumber\\
&&\nonumber\\
&=& \Omega^{-2}g^{ac} \left[ -2\Omega^{-3} j_c \nabla_a \Omega 
+\Omega^{-2} 
\nabla_a j_c -\Omega^{-3} \left( j_a \nabla_c \Omega +j_c \nabla_a \Omega 
- g_{ac} j_d \nabla^d \Omega \right) \right] =0 \,,\nonumber\\
&&
\end{eqnarray}
where we used Eq.~(\ref{Gammascaling}) and the covariant conservation of 
the electric charge $\nabla^c j_c=0$ in the unrescaled spacetime.

\section{Wave equation for the four-potential}
\label{sec:3}

The second question to address is whether the wave equation 
satisfied by the four-potential $A^b$ is conformally invariant. Since 
$A^b$ 
couples explicitly to the Ricci tensor and, contrary to the Maxwell tensor 
$F_{ab}$, is gauge-variant, this question is not trivial. 

The Maxwell equation~(\ref{Maxwells2}) guarantees that the Maxwell tensor 
can be  derived from a four-potential $A^b$ according to
\be
F_{ab}=\nabla_a A_b - \nabla_b A_a 
=\partial_a A_b - \partial_b A_a 
\,. \label{defpotential}
\ee
We will now go over the standard derivation of the equation satisfied by $A^b$ in curved spacetime. However, contrary to common practice, let us allow for the presence of sources. Furthermore, we won't fix the gauge in order to keep the presentation general.

In conjunction with Eq.~(\ref{defpotential}), the Maxwell 
equation~(\ref{Maxwells1}) yields 
\be
\Box A_b -\nabla^a\nabla_b A_a =-4\pi j_b \,. \label{questa}
\ee
Using the rule for the commutator of covariant derivatives in 
curved spacetime \cite{Wald, Carroll}
\be
\left[ \nabla_a , \nabla_b \right] A_d = {R_{abd}}^e A_e 
\ee
and the symmetries of the Riemann tensor to write 
\be
{R^c}_{bcd}=- { {R_b}^c}_{cd} = -{R_{dcb}}^c \equiv R_{db} \,,
\ee
the second term on the left hand side of Eq.~(\ref{questa})  becomes
\be
\nabla^c \nabla_b A_c = \nabla_b  \nabla^c A_c  + R_{db} A^d \,,
\ee
so that  (Ref.~\cite{MTW}, p. 569)
\be
\Box A_b -\nabla_b  \nabla^c A_c  -{R_b}^d A_d = -4\pi j_b 
\,. \label{eq:eqpotential}
\ee
This equation simplifies in the Lorentz gauge $\nabla^c A_c=0$, in which 
it is usually presented. Before investigating the conformal invariance of 
Eq.~(\ref{eq:eqpotential}), it is necessary to establish the scaling law 
of 
$A^b$. The validity of the second Maxwell equation in the conformally 
rescaled spacetime guarantees that the rescaled Maxwell tensor can be 
written as
\be
\tilde{F}_{ab}=\tilde{\nabla}_a  \tilde{A}_b - \tilde{\nabla}_b 
\tilde{A}_a 
= \partial_a  \tilde{A}_b - \partial_b \tilde{A}_a \,.
\ee
If the conformal weight of the four-potential is $p$, {\em 
i.e.}, $\tilde{A}_a = 
\Omega^p A_a$, then
\be
\tilde{F}_{ab} = p \, \Omega^{p-1} 
\left( \partial_a \Omega \, A_b - \partial_b \Omega \, A_a \right)
+\Omega^p F_{ab}  
\ee
and then the result $\tilde{F}_{ab}=F_{ab}$ gives 
\be
\left( \Omega^p -1 \right) F_{ab} +p \, \Omega^{p-1} \left( \partial_a 
\Omega \, 
A_b - \partial_b \Omega \, A_a  \right) =0\,.
\ee
This equation is only satisfied if $p=0$, or 
\be
\tilde{A}_b=A_b \,, \;\;\;\;
\tilde{A}^b= \Omega^{-2} A^b \,. \label{eq:transfA}
\ee
Having established this result, we can now proceed to check the conformal 
invariance of Eq.~(\ref{full}). First, one computes
\be
\tilde{\nabla}_c \tilde{A}_b = \nabla_c A_b -A_c \nabla_b \ln \Omega -A_b 
\nabla_c \ln\Omega +g_{bc} \, A_e \nabla^e\ln\Omega \,. \label{eq:35}
\ee
Before proceeding, we discuss gauge invariance.

\subsection{Lorentz gauge} 
\label{sec:3Lorentzgauge}

Our question about the conformal invariance of the Lorentz gauge can 
now be answered by computing
\be 
\tilde{g}^{ab} \tilde{\nabla}_a \tilde{A}_b = \Omega^{-2} \left( \nabla^a 
A_a +2A_a 
\nabla^a \ln \Omega \right) \,. \label{mah}
\ee
The Lorentz gauge is broken by the conformal transformation unless  
the gradient of the conformal 
factor is perpendicular to the four-potential, 
\be
A^c \nabla_c \Omega = 0 \,, \label{tildeLorentz}
\ee
a very special condition that cannot be enforced in general. 
Therefore, conformal transformations break the Lorentz gauge or, in 
general, any  gauge condition (this result appears in 
Ref.~\cite{SonegoFaraoniJMP} which applies it to the study of the sharp 
propagation of electromagnetic waves in special curved spacetimes). 
However, this is a ``soft'' breaking of conformal invariance that can 
always 
be removed by a gauge redefinition. If, in the absence of sources, one 
starts with the Lorentz gauge 
$\nabla^c A_c=0$, one ends with $\tilde{\nabla}^c \tilde{A}_c = 
2\Omega^{-3} A_b \nabla^b \Omega \neq 0 $, but it is always possible to 
perform a gauge transformation to restore the Lorentz gauge \cite{Wald}. 
As such, 
the term introduced in the equation for the four-potential by the 
conformal transformation can be gauged away and gives zero contributions 
to $\tilde{F}_{ab}$.

\subsection{Light-cone gauge}

To avoid the issue of having to fix the gauge every time one performs a 
conformal transformation to the system, it is possible to make a different 
choice for the gauge-fixing condition. Indeed, an adequate alternative 
to the 
Lorentz gauge, which is usually the go-to Lorentz-invariant condition 
used whenever one deals with the study of gauge theories in both 
classical and quantum field theory, is given by the so called {\em 
light-cone gauge}. Let $\ell ^a = \ell ^a (x^\alpha)$ be the tangent 
vector field to a congruence of null geodesics of the spacetime 
$(\mathcal{M}, g_{a b})$. Then the condition
\be 
\ell ^a A_a = 0
\ee
defines the light-cone gauge. Clearly, this condition is conformally  
invariant and therefore all conformally equivalent frames would agree on 
this gauge-fixing. However, this choice comes at a price, namely this 
condition depends on the choice of the congruence of null geodesics and 
its implications for the equations of motion are not as apparent as in 
the case of the Lorentz gauge.

\subsection{Conformal invariance of Eq.~(\ref{full})}

Continuing our calculation that led to Eq.~(\ref{eq:35}), one computes
\be
\tilde{\Box} \tilde{A}_b = \Omega^{-2} g^{ac} \tilde{\nabla}_a 
\tilde{\nabla}_c A_b 
\ee
where, using the transformation properties~(\ref{eq:t1})-(\ref{eq:t4}),  
one obtains
\begin{eqnarray}
\tilde{\nabla}_a \tilde{\nabla}_c \tilde{A}_b &=& \nabla_a \nabla_c A_b  - 
\nabla_b \ln \Omega \left( \nabla_a A_c +\nabla_c A_a \right) 
\nonumber\\
&&\nonumber\\
&\, & - A_c \nabla_a 
\nabla_b \ln \Omega -2 \nabla_a A_b \nabla_c \ln \Omega  
-A_b \nabla_a \nabla_c \ln \Omega  \nonumber\\
&&\nonumber\\
&\, & +g_{bc} \nabla_a \left( A_e \nabla^e \ln \Omega \right) -2 \nabla_c 
A_b \nabla_a \ln \Omega  +3 A_c \nabla_a \ln \Omega \nabla_b \ln \Omega  
+3A_b \nabla_a \ln \Omega \nabla_c \ln \Omega \nonumber\\
&&\nonumber\\
&\, &  -2g_{bc} \nabla_a \ln  \Omega \, A_e \nabla^e \ln \Omega  +2 A_a 
\nabla_b \ln \Omega \, \nabla_c \ln  \Omega \nonumber\\
&&\nonumber\\
&\, &+ g_{ab} \left( \nabla^e \ln \Omega \nabla_c A_e - A_c \nabla^e \ln 
\Omega \nabla_e \ln \Omega  -\nabla_c \ln\Omega \, A_e 
\nabla^e\ln\Omega \right) \nonumber\\
&&\nonumber\\
&\, & +g_{ac} \left( \nabla^e \ln \Omega \nabla_e A_b - A_b \nabla^e \ln
\Omega \, \nabla_e \ln \Omega - \nabla_b \ln \Omega \, A_e \nabla^e \ln 
\Omega \right) \,.
\end{eqnarray}
The contraction of this equation produces the d'Alembertian
\begin{eqnarray}
\tilde{\Box} \tilde{A}_b &=&  \Omega^{-2} \left[ 
\Box A_b -A_b \Box \ln \Omega -2 \nabla_b \ln \Omega \nabla^a A_a 
- A^a \nabla_a \nabla_b \ln \Omega 
+\nabla_b \left( A_a \nabla^a \ln \Omega \right) \right. \nonumber\\
&&\nonumber\\
&\, & \left. -2  A_a \nabla^a \ln \Omega \, \nabla_b \ln \Omega 
 -2 A_b \nabla^a \ln\Omega \, \nabla_a\ln\Omega +\nabla^a \ln\Omega 
\, \nabla_b A_a  \right]  \,,
\end{eqnarray}
while the $\tilde{R}_b^{\; d} \tilde{A}_d$ term of 
Eq.~(\ref{eq:eqpotential}) is
\begin{eqnarray}
\tilde{g}^{ad} \tilde{R}_{ab} \tilde{A}_d  &=& \Omega^{-2} \Big( {R_b}^d 
A_d -2 A^a \nabla_a \nabla_b 
\ln\Omega -A_b \Box\ln\Omega +2 A_a \nabla^a 
\ln\Omega \, \nabla_b \ln\Omega  \nonumber\\
&&\nonumber\\
&\, &  -2A_b \nabla^a  \ln\Omega \, \nabla_a \ln\Omega 
\Big) \,.
\end{eqnarray}
The covariant differentiation of Eq.~(\ref{mah}) gives  
\begin{eqnarray}
\tilde{\nabla}_b \tilde{\nabla}^a \tilde{A}_a &=& \nabla_b 
\tilde{\nabla}^a \tilde{A}_a 
= \Omega^{-2} \left[
\nabla_b \nabla^a A_a +2A^a \nabla_a \nabla_b \ln\Omega -2\nabla_b 
\ln\Omega \, \nabla^a A_a \right. \nonumber\\
&&\nonumber\\
&\, & \left. -4  A_a\nabla^a \ln\Omega \, \nabla_b \ln\Omega 
+2\nabla_b A_a \nabla^a \ln\Omega \right] \,.
\end{eqnarray}
Putting everything together and using the 
transformation law~(\ref{tildej}) of the four-current yields 
\be
\tilde{\Box} \tilde{A}_b - \tilde{\nabla}_b \tilde{\nabla}^a \tilde{A}_a 
-\tilde{R}_b^{\; d} \tilde{A}_d 
=-4\pi \tilde{j}_b \,, 
\ee
which demonstrates the conformal invariance of the full 
equation~(\ref{full}) satisfied by 
the four-potential prior to fixing the gauge. The Lorentz gauge version of 
this equation commonly reported in the literature is not conformally 
invariant 
because the Lorentz gauge is broken by the conformal transformation.

\section{Wave equation for the Maxwell field}
\label{sec:3bis}

The Maxwell field $F_{ab}$ also satisfies a wave equation. Let us consider 
the condition $\nabla _{[a} F_{bc]} = 0$; if we 
unpack it and differentiate both sides, we obtain
\be
\Box F_{ab} + \nabla ^c \nabla_b F_{ca} + \nabla ^c \nabla_a F_{bc} = 0 \, ,
\ee
or 
\be
\Box F_{ab} + g^{cd} \left( \nabla _d \nabla_b F_{ca} + \nabla _d \nabla_a 
F_{bc} \right) = 0 \, .
\ee
Using the identity
\be
\left[ \nabla_a , \nabla_b \right] {F}_{cd} =  R_{abce} {F^e}_d + 
R_{abde} {F_c}^e \,,\label{Ricciidentity}
\ee
one finds
\be
\nabla _d \nabla_b F_{ca} + \nabla _d \nabla_a F_{bc} =
\nabla _b \nabla_d F_{ca} + \nabla _a \nabla_d F_{bc} +
R_{dbce} {F^e}_{a} + R_{dbae} {F_c}^{e} +
R_{dabe} {F^e}_{c} + R_{dace} {F_b}^{e}
\ee
that, contracting with $g^{cd}$, gives
\be
g^{cd} \left( \nabla _d \nabla_b F_{ca} + \nabla _d \nabla_a F_{bc} \right) =
\nabla _b \nabla ^c F_{ca} + \nabla _a \nabla ^c F_{bc} +
(R_{cabe} - R_{cbae}) F^{ec} + 
{R^c}_{bce} {F^e}_a + {R^c}_{ace} {F_b}^e \, .
\ee
Now, recalling \eqref{Maxwells1}, that ${R_{abc}}^b \equiv R_{ac}$, and 
that $R_{a[bcd]}=0$, together with the symmetry properties of the Riemann 
tensor, one obtains \cite{Tsagas, StarkoCraig}
\be
\Box F_{ab} + R_{abcd} F^{cd} -R_{ac} {F^c}_b +R_{bc} {F^c}_a = 4\pi 
\left( \nabla_b j_a -\nabla_a j_b \right) \,. \label{eq:ruppa}
\ee
Since this equation is derived using only the 
identity~(\ref{Ricciidentity}) and the Maxwell equations, which have 
already been shown to be conformally invariant, together with the 
symmetries of the Riemann tensor (also invariant), we only need to show 
the conformal invariance of Eq.~(\ref{Ricciidentity}) to establish the 
conformal invariance of the wave equation~(\ref{eq:ruppa}). But 
Eq.~(\ref{Ricciidentity}) is clearly conformally invariant
because it derives from $\nabla _{[a} F_{bc]} = 0$, which is  
equivalent to $\partial _{[a} F_{bc]} = 0$ and is conformally invariant.

\section{Conclusions}
\label{sec:4}

There remains to check the consistency of the various scaling laws 
derived above with the conformal invariance of the 
action~(\ref{emaction}), but this is easy to do. Using the transformation 
properties (\ref{conftransf}), (\ref{eq:t1}), (\ref{eq:t2}), 
(\ref{eq:20}), (\ref{eq:transfA}), and (\ref{tildej}), one obtains
\begin{eqnarray}
\sqrt{-g} \, g^{ac} g^{bd} F_{ab}F_{cd} &=&  \Omega^{-4} \sqrt{-\tilde{g}} 
\, \Omega^2 \tilde{g}^{ac} \, \Omega^2 \tilde{g}^{bd} \tilde{F}_{ab} 
\tilde{F}_{cd} \nonumber\\
&&\nonumber\\
&=& \sqrt{-\tilde{g}} \, \tilde{g}^{ac}  \tilde{g}^{bd}  \tilde{F}_{ab} 
\tilde{F}_{cd} 
\equiv  \sqrt{-\tilde{g}} \, \widetilde{  \tilde{F}_{ab} \tilde{F}^{ab} } 
\,,\\
&&\nonumber\\
\sqrt{-g} \, g^{ab} A_a j_b &=& 
\Omega^{-4} \sqrt{-\tilde{g}} \, \Omega^{2} \tilde{g}^{ab}  \tilde{A}_a 
\, \Omega^{2} \tilde{j}_b = \sqrt{-\tilde{g}} \, \tilde{g}^{ab}  
\tilde{A}_a \tilde{j}_b \equiv  \sqrt{-\tilde{g}} \, \widetilde{ 
\tilde{A}^c  \tilde{j}_c} \,.
\end{eqnarray}
Putting everything together, the Maxwell action~(\ref{emaction}) becomes
\be 
S_{(em)} = \int d^4 x \sqrt{-g} \, \left( -\frac{1}{4} \, F_{ab}F^{ab} 
+4\pi A_b j^b \right) = 
\int d^4 x \sqrt{-\tilde{g}} \, \left( -\frac{1}{4} \, 
\widetilde{ \tilde{F}_{ab} \tilde{F}^{ab} } 
+ 4\pi \widetilde{ \tilde{A}_b \tilde{j} ^b } \right) \,, 
\ee
{\em i.e.}, it is invariant in form under the conformal 
transformation~(\ref{conftransf}). Hence, it produces conformally 
invariant field equations provided that $F_{ab}, A^c$, and $j^c$ 
transform according to the rules discussed in the previous sections.

Using the conformal scaling laws discussed, we derive immediately the 
transformation property of the Maxwell stress-energy tensor~(\ref{Tab}) 
\be
\tilde{T}_{ab} = \Omega^{-2} T_{ab} \,, \;\;\;\; 
\tilde{T}^{ab}=\Omega^{-6} T^{ab} \,, \;\;\;\;\;
\tilde{T}_a^{\; b}= \Omega^{-4} {T_a}^b 
\ee
and, of course, $\tilde{T}=\Omega^{-4} T=0$ \cite{Bekenstein}. This 
completes the analysis 
of the conformal invariance of electromagnetism in curved spacetime. In 
the geometric optics approximation, electromagnetic waves satisfying the 
Maxwell equations follow null rays, which obey the null geodesic equation
\be
\frac{d^2 x^\alpha}{d\lambda^2}+\Gamma^{\alpha}_{\beta \gamma} 
\, \frac{dx^\beta}{d\lambda} \frac{dx^\gamma}{d\lambda}=  0 \,,
\ee
where $\lambda$ is an affine parameter along the geodesic and $k^\alpha 
=dx^\alpha/d\lambda$ is the four-tangent to the null geodesic. As a 
consequence of the conformal invariance of the Maxwell equations, 
conformal transformations leave null geodesics 
invariant (apart from changing the parametrization to a non-affine 
parameter), a  result that could also be established directly without 
knowledge of the conformal invariance of Maxwell's theory \cite{Wald, 
Carroll}.

Conformally invariant systems are rare in nature and the electromagnetic 
interaction realizing this invariance is studied intensely because it is 
one of only four fundamental forces and the simplest example of gauge 
theory.

\section*{Acknowledgments}

This work is supported, in part, by the Natural Science and Engineering 
Research Council of Canada (Grant No. 2016-03803 to V.F.) and by Bishop's 
University.

\end{document}